# Adversarial Example Attacks against ASR Systems: An Overview

Xiao Zhan[a,b], Hao Tan[a,b], Xuan Huang[c,d], Denghui Zhang[a], Keke Tang[a], Zhaoquan Gu[a,b,e]

[a] Guangzhou University, Guangzhou, China
[b] Department of New Networks, Peng Cheng Laboratory, Shenzhen, China
[c] The 5th Electronic Research Institute, Ministry of Industry and Information Technology, Guangzhou, China
[d] Laboratory of MIIT for Intelligent Products Testing and Reliability, Guangzhou, China
[e] National University of Defense Technology, Changsha, China
guinianzx@gmail.com, th198@e.gzhu.edu.cn, huangxuan@ceprei.biz, {denghui.zhang, kktang, zqgu}@gzhu.edu.cn

*Abstract*—With the development of hardware and algorithms, ASR(Automatic Speech Recognition) systems evolve a lot. As The models get simpler, the difficulty of development and deployment become easier, ASR systems are getting closer to our life. On the one hand, we often use APPs or APIs of ASR to generate subtitles and record meetings. On the other hand, smart speaker and self-driving car rely on ASR systems to control AIoT devices. In past few years, there are a lot of works on adversarial examples attacks against ASR systems. By adding a small perturbation to the waveforms, the recognition results make a big difference. In this paper, we describe the development of ASR system, different assumptions of attacks, and how to evaluate these attacks. Next, we introduce the current works on adversarial examples attacks from two attack assumptions: white-box attack and black-box attack. Different from other surveys, we pay more attention to which layer they perturb waveforms in ASR system, the relationship between these attacks, and their implementation methods. We focus on the effect of their works.

## I. INTRODUCTION

ASR(Automatic Speech Recognition) refers to converting Speech signals(such as waveform) into text. In various machine learning technics, ASR belongs to NLP(Natural Language Processing). Unlike SVM, Linear Regression, K-Means, CNN, RNN, Transformer, it is not an algorithm. ASR is a system. It contains audio processing like VAD, data cleaning, data augmentation, acoustic model, language model, etc.

In 1970s, people still regarded ASR impossible. Things changed in 1990s with the introduction of the HMM-GMM model. CMU's Kai-Fu Lee, et al developed the Sphinx open source speech recognition toolkit and the Cambridge University team developed the HTK(Hidden Markov Model Toolkit) toolkit. These toolkits helped people develop tranditional ASR systems. In 2010, as deep learning became a hotspot in academia and industry, HMM-DNN quickly replaced the HMM-GMM, cause DNN fits better than GMM as acoustic model. JHU's Daniel Povey led the development of Kaldi, a next-generation speech recognition toolkit at this time. The recognition accuracy of such hybrid ASR systems supported by Kaldi had surpassed that of human minute-takers. With the popularity and demand of end-to-end neural networks around 2020, ESPnet, WeNet came into being, which support powerful end-to-end speech recognition models, such as CTC, Transformer, etc. Nowadays, ASR has come to everyone. On AIoT devices, we can use speech recognition to control various functions of self-driving cars; use it to control smart speakers and thus various home appliances. On terminal devices, we can use ASR to recognize the music playing in the current environment, recognize foreign speakers on Youtube we don't understand and then automatically translate them, etc.

In recent years, there are many research works on the vulnerability of DNNs in image recognition. Speech recognition also faces serious security problems. As it is difficult to achieve complete robustness, ASR is easy to get wrong answer due to a small perturbation. Moreover, since ASR is very close to our life, there will be much more dangerous security problems when attack happens. Even personal safety is at risk.

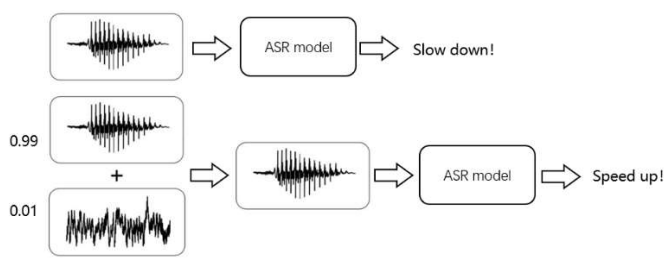

Figure 1: an instance of adversarial example against ASR

There are some recent research works on advesarial example attacks against ASR systems. Researchers mainly focus on black-box and white-box attacks. On white-box attacks, they work with acoustic model structures. They want to generate adversarial examples for different ASR systems, making adversarial examples targeted/non-targeted, universal, imperceptible, etc. On black-box attacks, researchers initially migrate the adversarial examples from white-box attacks to black-box attacks, and later use alternative models, evolutionary algorithms, etc. to generate adversarial examples. Beyond these, some people directly started from the data layer to construct adversarial examples by exploiting the psychological masking of the human ear in frequency domain. However, the existing works do not have a good summary, and the few reviews of adversarial attacks against ASR are not comprehensive and systematic enough.

In this paper, we summarize the progress and works on adversarial example attacks against ASR systems. First, we introduce various models of speech recognition in a chronological order, followed by background of adversarial attacks and evaluation metrics of the attacks against ASR systems. Then we analyze the current progress of adversarial example attacks against ASR systems in terms of both white-box and black-box. On the one hand, this paper is more systematic in presenting works in adversarial attacks against ASR systems and focuses more on the background of these works, the targets and results of the attacks, and the relationship between them. On the other hand, we refer to the latest works.

The outline of this paper is as follows: 1. introduce necessary of work on adversarial example attacks against ASR systems; 2. introduce the featurization and models involved in ASR systems; 3. introduce the background knowledge of adversarial example attacks against ASR systems; 4. formulate the current works; 5. conclude.

## II. BACKGROUND OF ASR

This section will introduce the background of speech recognition, starting with featurization and then introduce various popular speech recognition models involved in ASR systems from the past to the present.

### A. Featurization

First step is pre-emphasis. Because the energy of the high frequency part decays fast, so it is necessary to increase the energy of the high frequency part. Next is framing. As speech signal has the feature of short-time invariance, waveform is divided into one 25ms frame, so that it can represent the short-time signal characteristics. Third step is to make product of window function and each frame to prevent the generation of frequency leakage. The fourth step is Fourier transform, the time spectrum is replaced by the frequency spectrum. The fifth step is Mel filter set based on human auditory perception to obtain Filter bank features. The last step is to perform IDFT transform to obtain MFCC features.

### B. Models in ASR

*1) HMM-GMM*: In the speech recognition task, our goal is transform sequence X of audio signals into sequence Y of text as shown in Equation 2.1:

$$Y^* = \arg\max_Y P(Y \mid X) \quad (2.1)$$

We can rewrite the above equation using the Bayesian formula as:

$$\arg\max_Y \frac{P(X \mid Y)P(Y)}{P(X)} \quad (2.2)$$

P(X) equals 1. So it is only necessary to calculate P(X|Y). This former likelihood probability is the acoustic model often referred to in ASR systems. And this latter prior probability is the language model.

HMM-GMM is such a generative model. It uses HMM-GMM to calculate the probabilities of acoustic models, and the probabilities of language models are calculated with N-gram additionally. It uses the modeling units of the language model (e.g., triphoneme, phoneme, word) as the hidden variables of the HMM and the input feature sequences (e.g., MFCC, Fbank) as the observed variables. The emission probabilities from the hidden variables to the observed variables are modeled using GMM. The HMM-GMM is trained using the Viterbi algorithm, and the parameters of the GMM are updated using the EM algorithm.

*2) HMM-DNN:* With the popularity of deep learning, people are gradually incorporating deep learning into ASR. Among the various schemes, the most influential one model is HMM-DNN. After training HMM-GMM, we already know which observed variables belong to which hidden variables. Then we can use DNN to classify the features as different language model modeling units (referred to by state later) to obtain P(a|x). Once again, this posterior probability can be turned into the likelihood probability that we need to calculate from hidden variables to observed variables using the Bayesian formula. The transformation process is shown in Equation 2.3:

$$P(x \mid a) = \frac{P(x,a)}{P(a)} = \frac{P(a \mid x)P(x)}{P(a)} \quad (2.3)$$

*3) CTC:* HMM-based models effectively use HMM to build a topology of speech features to states so as to align them and model emission probabilities using other models. And with the advent of the deep learning, it is finally possible to replace the HMM as well. The acoustic model of ASR is constructed directly using Encoder+CTC Decoder using CTC loss. Also, this kind of model computes the posterior probabilities directly without transformations like the previous generative models.

CTC is specifically topologized by removing consecutive repetitive text units and then removing the null flag. Although CTC already computes the posterior probabilities of feature sequences to text sequences directly, they are often used in conjunction with language models, which tend to enhance the generalization performance of the whole system for specific purposes. The output process of CTC is shown in Figure 2.

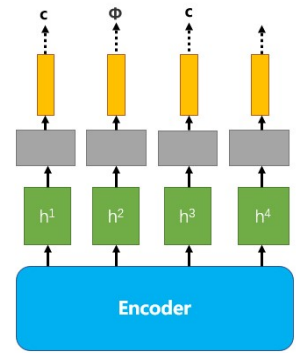

Figure 2: CTC based ASR model

*4) RNN-T:* CTC is the first end-to-end acoustic model rather than traditional model. Without a lengthy iterative training process, it greatly facilitates the development of ASR systems. But it still has its own shortcomings. It does not use the temporal information, just input a feature and output a state. People first try adding temporal model RNN to encoder. Then T'time step output will be added to T + 1'time input, no longer need to remove the continuous repetition of the output. This is RNA model. From RNA to RNN-T, some improvements are made: 1. input a feature can output more than one state; 2. language model is introduced. The output of T'time is obtained from the output of language model layer, and if the output of T+1'time is encountered as the empty flag, the memory information input to T+2'time is still the output obtained at T'time.

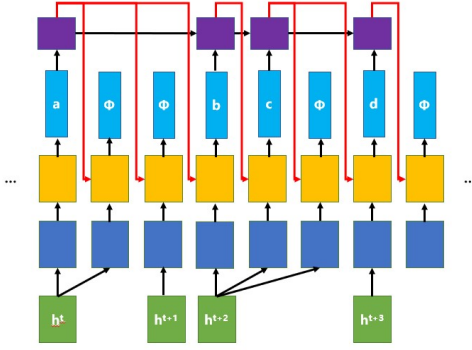

Figure 3: RNN-T based ASR model

*5) LAS:* LAS is a migration of the Seq2Seq model from NLP, using Encoder-Decoder structure and aligning with cross Attention. In decoding, tricks such as beam search, Teacher Forcing, and Location-aware attention are used to help train LAS.

*6) Transformer / Conformer:* Transformer is just use Transformer to solve ASR. Input the voice and output the text. Conformer is a combination of Transformer and CNN. It is known to all thar Transformer is good at modeling the original sentence context due to its self-attentive mechanism, but is less capable of extracting fine-grained local features. While CNN makes effective use of local features. Combining these two network structures can take into account both global interactions and local relevance. This end-to-end network structure is the most popular and advanced speech recognition network structure in 2022.

*7) The Latest ASR Pipeline:* **In the latest ASR systems**, we generally need 1. data cleaning, 2. data enhancement, 3. train acoustic models (sometimes combined with pre-trained models, such as Wav2Vec, etc.), 4. train language models 5. design decoding frameworks after the training.

III. BACKGROUND OF ADVERSARIAL ATTACK AGAINST ASR

This section introduces the background knowledge of adversarial example attacks against ASR systems, including the workflow of these attacks, the classification of attack assumptions, and the metrics of attack effectiveness.

*A. Theory of Adversarial Attack*

Adversarial example attacks refer to a security threat to a machine learning model caused by insufficient robustness due to the training effect not reaching the ideal state. As shown in the figure 4, specifically, an adversarial example attack perturbs a example that would be fed into a machine learning model to output normal results, making it to cross the decision boundary and letting the model output the wrong results.

As for adversarial example attacks against, it is perturbing the audio input to the ASR system so that the ASR system makes an error in the output recognition result of this audio. In addition, we often need to make it imperceptible for people to hear the change in the audio, so that they do not know themselves under attack.

In this process, sometimes we are not concerned with the outcome of the perturbation, and sometimes the perturbation needs to be restricted. Also, there are many classifications for attack assumptions: targeted/untargeted attacks; target of the perturbation; universal perturbation/specific perturbation; WTA attacks/WAA attacks; black-box attack/white-box attack, etc.

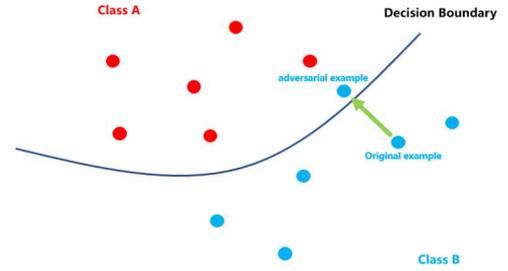

Figure 4: theory of adversarial attack

*B. Non-Targeted Attack / Targeted Attack*

The Non-targeted attack means to make ASR system output an wrong result, without making it output a specific result. As shown in Equation 3.1. x represents original audio, $\Delta$ is the perturbation. Function F() stands for ASR system.

$$F(x + \Delta) != F(x) \quad (3.1)$$

Targeted attack means to make ASR system output a wrong and specific result. For an example, we need to make an audio signal that the ASR system originally recognized as "Accelerate!" recognized as "Decelerate!" after perturbing the audio. As shown in Equation 3.2:

$$F(x) = A, \ F(x + \Delta) = B \quad (3.2)$$

*C. Perturbation Objective*

The objects of perturbation can also be various. It can be audio's waveforms or audio's features. If the object of perturbation is a feature. Then after obtaining the adversarial example of the feature, the waveform needs to be synthesized from the feature in reverse. While the object of perturbation is a waveform, there is no need to synthesize it in reverse. Regarding the choice of the object of perturbation, if the input of the model is a feature, then the object of perturbation is the feature, and if the input of the model is audio, then the object of perturbation is the audio.

*D. Universal Perturbation / Specific Perturbation*

A specific perturbation means that the generated perturbation can only cause a specific audio to be recognized incorrectly by the ASR system.

A universal perturbation means that the generated perturbation combined with any audio will cause the audio to be incorrectly recognized.

$$F(x \text{ in } \mu \text{ set} + \Delta) != F(x) \tag{3.3}$$

*E. WTA Attack / WAA Attack*

WTA attack refers to WAV-TO-API attack and WAA attack refers to WAV-AIR-API attack. As the name implies, WTA attack refers to the attack on the API of the machine learning system. The adversarial example is directly input into the API of the speech recognition system, such as directly inputting the adversarial example into the locally running own ASR system on the computer, or inputting the ASR API provided by the service provider. While WAA attack refers to the adversarial attack on the speech recognition system of the actual physical environment attack, such as playing an adversarial example with speaker A, and then receiving the voice with receiver B a few meters away, and inputting the voice signal into ASR system. In the WAA attack, factors such as reverberation, echo, and white noise need to be considered.

*F. White-Box attacks / Black-Box attacks*

White-box attack refers to model structure, parameters, and superparameters of the target ASR system are known. Black-box attack means that the internal structure of the target ASR system is unknown, and only queries can be launched against it to return the recognition result or topK confidence level.

*G. Metrics*

In attacks against ASR system, there are three relatively important metrics: WER, SNR, and success rate. WER refers to the word error rate and is calculated as follows:

$$WER = \frac{S + D + I}{N} \tag{3.4}$$

Where N represents the number of words in the sentence, S, D, and I respectively represent the words incorrectly inserted, incorrectly deleted, and incorrectly substituted in the recognition result. In attack, we will define a WER threshold value. Only when WER is bigger than the threshold, the attack is regarded success. SNR refers to the signal-to-noise ratio, which is the ratio of Power of Signal to Power of Noise. As shown in Equation 3.5:

$$\text{SNR (dB)} = 10 \log_{10}\left(\frac{P_{\text{signal}}}{P_{\text{noise}}}\right) = 20 \log_{10}\left(\frac{A_{\text{signal}}}{A_{\text{noise}}}\right) \tag{3.5}$$

And success rate refers to the success rate of the adversarial examples generated with this method against the target model in a specific attack assumption.

In addition, there are some other metrics. For example, in WAA, there is also a metric of the number of attacks, because each WAA attack experiment is not stable and subject to various influences. So even attacked successfully, there exist some adversarial examples that are attacked 100 times with 1 success and some adversarial examples that are attacked 100 times with 10 successes.

## IV. METHODS OF ADVERSARIAL ATTACK AGAINST ASR SYSTEM

The following describes the current research progress of adversarial example attacks against ASR in terms of both white-box attacks and black-box attacks.

*A. White-Box Attacks*

In white-box adversarial attacks, the focus of research is generally on the generation of different adversarial examples or the tricks used by various researchers to make the adversarial attacks more robust and imperceptible, often using algorithms such as FGSM, FGM, I-FGSM, MI-FGSM, CW, etc. In this section, we will not introduce each of these white-box adversarial methods, but focus on the state of the art of adversarial attack research by researchers in the white-box attacks against ASR systems, including the features, mothohs, frameworks and models of their attacks, and the metrics.

In 2017, Gong et al.[12] generated adversarial examples against ASR system based on a End2End model for the first time. They used FGSM to generate the adversarial examples, while the object of perturbation is directly the raw waveform. So there is no lossy conversion between the feature sequence and the speech audio. Here are their optimization objectives and the FGSM optimization function:

$$\text{Minimize } \|\eta\| \text{ s.t. } f(\mathbf{x} + \eta) \neq f(\mathbf{x})$$
$$\eta = \epsilon \text{sig n}(\nabla_{\mathbf{x}} J(\theta, \mathbf{x}, y)) \tag{4.1}$$

They used a WaveCNN model modified from the popular WaveRNN model at that time in order to achieve direct perturbation of the raw audio. Finally they succeeded to make the success rate of untargeted attacks high enouch with small perturbations. In the same year, Cisse et al.[13] used their proposed Houdini method to generate adversarial examples and tested them on DeepSpeech2, achieving a similar success rate. Because their works is only non-targeted, so it is not very threatening.

In 2018, Yuan et al.[14] performed a targeted attack by migrating CW, an optimization-based approach, to adversarial example attacks against ASR system. They attacked the ASPIRE CHAIN MODEL in the Kaldi framework, which is an HMM-DNN model. Specifically, they attacked the feature extraction part and the DNN part of the model, i.e., they perturbed the audio before the input feature extraction and optimized the pdf-ids of the DNN output. They repeatedly embed the perturbation into a section of the song so that it only needs to be successfully attacked once, and added random noise so that the success rate of the WAA attack will be greatly increased. The following equation 4.2 is the optimization function they used to generate the adversarial examples. g(x) represents the input audio to the DNN output pdf-ids of this model, μ(t) represents the perturbation, n(t) represents the random noise, b represents the target instruction output pdf-ids. To make the result of the perturbed audio as close to the original as possible, their optimization function uses the $l_1$ parametrization, the perturbation The limits on the size of the perturbation use the $l_\infty$ parametrization. They finally achieve a successful attack with a very low SNR of 1.3-1.7 dB.

$$\arg\min_{\mu(t)} \| g(x(t) + \mu(t) + n(t)) - \mathbf{b} \|_1$$
$$|\delta(t)| \leq l, n(t) = \operatorname{rand}(t), |n(t)| \leq N \quad (4.2)$$

In 2018, Carlini et al.[15] also performed targeted attacks, while he attacked DeepSpeech, which is LSTM-based neural network using CTC loss, and the target of perturbation is raw waveform. The function they optimize in generating the adversarial examples is based on CTC, directly target labeled under loss, while the control of perturbation size uses $l_2$ paradigm to avoid the problem of oscillation non-convergence generated by using $l_\infty$ paradigm. They use the following optimization functions:

$$\text{minimize } |\delta|_2^2 + c \cdot \ell(x + \delta, t)$$
$$\text{such that } dB_x(\delta) \leq \tau \quad (4.3)$$

Finally the SNR of their scrambled speech signal was 15dB to 45dB, with a high success rate of the attack. Based on their work, Das et al.[16] made ADAGIO, an online system capable of interactively generating speech recognition adversarial examples, which can recognize the audio uploaded by the uploader as text, and then the attacker can select "joanna" in the recognition result and replace it with "marissa "The system will return the generated adversarial audio. Yuan and Carlini's works have practical significance, but the attacked ASR systems are not a streaming ASR system and mainly attacked under WTA circumstances, so further research is needed.

Unlike controlling the size of perturbations on images to make them imperceptible, adversarial examples against ASR system can also control perturbations from the perspective of human perception. 2018 Schönherr et al.[17] used a psychoacoustics-based approach to produce adversarial examples. They took inspiration from the mp3 compression algorithm and used the original audio to calculate the hearing thresholds H and then limited the difference between the original signal spectrum S and the modified signal spectrum M to within the human perceptual threshold. Their experiments are based on Kaldi's WSJ example, based on the HMM-DNN model, while they integrated the preprocessing process into the DNN to facilitate backpropagation of the original audio adversarial examples.

Based on Schönherr's work, in 2019 Qin et al.[18] changed the model of the attack target to Lingvo and changed the generated adversarial example from a phrase to a full sentence. A smaller perturbation was first obtained using Carlini's method, and then the adversarial example was made imperceptible based on psychoacoustics. In 2020, Schönherr et al.[19] again combined RIR (Room Impulse Response) and psychoacoustics by adding a convolutional layer that mimics RIR in front of the feature extraction layer to simulate an actual WAA attack, and the experiments generated more robust adversarial examples based on Lingvo's LAS model.

In 2019, Liu et al.[19] proposed Sampling Perturbation Technology (SPT) and Weighted Perturbation Technology (WPT) to improve the efficiency of generating adversarial examples and the robustness of adversarial examples. SPT refers to fixing a part of the signals in the example and reducing the number of perturbed signals. They successfully reduced the number of perturbed signals to 5% of the original, which speeds up the generation of the adversarial examples. And WPT is to adjust the distortion weights at different locations to speed up the generation of the adversarial examples. They focused on adjusting the part of the audio corresponding to the different parts of the original label and target, and compared the difference in the effect of each formula for limiting the perturbation. Their experimental attack is on Deepspeech.

In addition to several perspectives of improving the robustness of adversarial examples and reducing human perception to do research on adversarial example attacks ASR systems, universal adversarial perturbations is also one hot spots of research. In 2019, Neekhara et al.[21] used their work to demonstrate that universal adversarial perturbations exist not only in adversarial example attacks against image tasks, but also in adversarial example attacks against ASR systems. The following are the targets of their attacks:

$$CER\big(C(x), C(x+v)\big) > t \text{ for "most" } x \in \mu \quad (4.4)$$

They need to find a perturbation that makes the CER or EditDistance large enough, and they set t=0.5,i.e., the CER should be more than 50%. The following is their optimization function, where the size of the perturbation uses the $l_\infty$ parametrization:

1. $\| v \|_\infty < \epsilon$
2. $P_{x \sim \mu}\big(CER(C(x), C(x+v) > t)\big) \geq \delta \quad (4.5)$

They initialized a perturbation V and updated V by iterating through the speech signals $X_i$ in the speech examples distribution μ. Specifically, they calculated the superimposed CER of V for each $X_i$ in each iteration. If the CER is less than a preset t, they need to find the smallest perturbation $\Delta V_i$ corresponding to $X_i$, and then update V with $\Delta V_i$ until the total attack success rate (CER proportion of speech signals greater than t) stops iterating when it reaches the preset value. And the method they used to find the minimum perturbation $\Delta V_i$ corresponding to $X_i$ is I-FGSM to do gradient descent on CTC loss. In each iteration their optimization scheme is as follows:

$$\text{Minimize } J(r) \text{ where}$$
$$J(r) = c \| r \|^2 + L\big(x_i + v + r, C(x_i)\big)$$
$$\text{s.t. } \| v + r \|_\infty < \epsilon$$
$$\text{where } L(x, y) = -\text{CTCLoss}(f(x), y) \quad (4.6)$$

In 2020, Zhuohang Li et al.[22] focused on the increasingly widely used streaming ASR systems. They argued that the universal adversarial perturbation in streaming ASR systems that does not need to be synchronized with the carrier audio is the more practical attack. So they researched on universal's adversarial examples for streaming ASR systems. They attacked keyword speech recognition based on convolutional neural networks, using the same loss function as the previous optimization-based loss function. And they need to generate sub-second perturbations so that they do not need to synchronize with the carrier audio. Their method for generating sub-second perturbations is as follows. r is the time delay, where the loss is computed with the perturbations superimposed on different time delays, so that the perturbations can be added anywhere in the streaming ASR system. x is the carrier audio, where the universal adversarial example is obtained by traversing the audios in the whole μ.

In 2021, Zhiyun Lu et al.[23] explored targeted universal adversarial perturbations for the now more popular ASR models of E2E, including LAS, CTC, and RNN-T. They used Additive perturbation and prepending perturbation to test the robustness of LAS, CTC, and RNN-T. It is found that LAS has the highest attack success rate, while CTC is more robust and less likely to be attacked successfully.

## B. Black-Box Attacks

Compared with white-box attacks, black-box attacks are more practical and involve more machine learning cross-field knowledge rather than solely ASR knowledge. So I will focus more on explaining the attack methods researchers used and the attack effects rather than the specific function formulas used. The following will now be divided into 4 categories of black-box attacks methods to explain.

*1) Featurization based Black-Box attacks*

In 2015, Vaidya et al.[24] proposed a method to adversarial example attacks at the data layer. They used TTS to generate the audios needed to get the target model to recognize correctly, then converted the audios to MFCC features and resynthesized the audios. By fine-tuning the MFCC features during this process, the goal was to make the ASR system successfully recognize but make it imperceptible to humans. Later in 2016 Carlini et al.[25] improved his work by introducing more realistic circumstances and noise after resynthesizing the audios, but the results achieved were not very good: the signal-to-noise ratio was only 15 dB and the attack results were very unstable. Although in most cases it succeeded to make the adversarial examples imperceptible to humans, it also made the accuracy of machine recognition much lower.

And in 2017, Zhang et al.[26] exploited the vulnerability that some ASR systems do not downsample high-frequency signals for DolphinAttack. They used the frequency domain portion of ultrasound to load instructions that needed to be recognized by the target ASR system, which not only allows the attack success rate to be effectively increased but also enables the human to completely hear no changes in the audios. However, since this kind of attack requires the target ASR system to be able to receive and recognize the identified high-frequency speech signal, it is extremely easy to make corresponding defense and difficult to replicate the attack. In 2018, in order to make the high-frequency adversarial attack more distant, Roy et al.[27] split the parts of the audio through multiple speakers, thus making the low-frequency leakage on their hardware more controllable.

In 2019, Abdullah et al.[28] proposed four methods to generate adversarial examples in the featurization process: Time Domain Inversion (TDI), Random Phase Generation (RPG), High-Frequency Addition (HFA), Time Scaling (TS ). TDI is interference in the time domain while leaving the frequency domain unchanged; RPG is a change in phase without changing the amplitude of the signal; HFA refers to the addition of high-frequency signals removed in signal processing to the audio, potentially masking instructions in the audio if the sine waveform is strong enough; TS refers to accelerating speech to the point where it can still be recognized by the machine while making it more difficult for a person to perceive instructions. Later in 2021, Abdullah et al.[29] proposed to use the DFT and SSA transforms respectively, and then restructured the audio after adjusting the features based on a pre-set threshold. Their attacks have strong randomness and are not easy to be defended. But the success rate of adversarial attacks is low in the process of feature transformation.

*2) Migration based Black-Box attacks*

Migration-based attacks refer to directly using the adversarial examples generated by white-box attacks on other black-box attacks. The effectiveness of this approach depends on the similarity between the white-box and black-box models, the generalization performance of the adversarial examples, etc. In Yuan's commandersong attack[14], adversarial examples were directly migrated to the black-box attack after the white-box attack were done. The success rate of such attacks is highly erratic and has a very low success rate in most circumstances.

*3) Substitute model based Black-Box attacks*

In 2020, Yuxuan Chen et al.[30] proposed Devil's Whisper. It is to construct a white-box ASR system, and to tune the white-box model by querying the black-box model and getting results. So that the adversarial examples generated from the white-box model are more robust to the black-box model. They took Kaldi's ASpIRE Chain model as the base model and trained a substitute model with local information using a small amount of data set obtained by querying the black-box. Then they updated the adversarial examples with the base model, and updated the adversarial examples locally with the substitute model. During this period, they used some tricks, such as using MI-FGM to avoid getting into local optimum, reducing the number of queries by dynamically adjusting the number of iterations of queries to the black-box model by their special settings, etc. In 2020, Fan et al.[31] used the method differential evolution in the process of generating adversarial examples using alternative models based on the construction of alternative models. In 2021, Chen Ma et al.[32] proposed the use of meta-learning models as alternative models to generate adversarial examples. Although they tried to minimize the number of queries to the black-box ASR systems when using the substitute model based black-box attacks. It still reached tens of thousands of queries and was easy to detect.

*4) Evolutionary Algorithm Based Black-box Attack*

In 2018, Alzantot et al.[33] proposed to use a genetic algorithm to generate adversarial examples without the need to use gradient descent to generate the adversarial examples. To minimize the effect of noise on human perception, they added noise only at the lowest valid bits of the waveform. If there is no successful attack and the maximum number of iterations is not reached, the examples are selected by the fitness score of the previous generation examples. The fitness scoring here means that the black-box ASR system will give scores to the TopK results of obtained adversarial examples. Then the next generation of adversarial examples are obtained by crossover and variation. Finally, 87% success rate of the attacks was achieved. Meanwhile people's perception of the adversarial examples was not high, 89% still identified the adversarial examples as the original audios. In the same 2018, Khare et al.[34] used evolutionary algorithms to attack the ASR model in Deepspeech and Kaldi framework, and in the targeted attack they used an optimization algorithm similar to white-box attacks, where the scoring of the adversarial examples do not only rely on the WER of the adversarial examples, but also takes into account the acoustic similarity with the original audios. Taori et al. 2019[35] used Momentum Mutation in this process to avoid falling into a local optimum during the iteration of the evolutionary algorithm.

In 2020, Du et al.[36] used PSO algorithm to generate adversarial examples based on the Deepspeech framework. They used the PSO algorithm to update the adversarial examples and modified the standard PSO algorithm to retain the current best particle in all iterations. In the case of a black-box attack, the PSO algorithm result is the final adversarial example; in the case of a white-box attack, they used optimization algorithms to fine-tune the results.

In 2021, Zheng et al.[37] proposed to use general cooperative co-evolution combined with CMA-ES evolutionary algorithm to generate audio adversarial examples in WTA attack scenarios. In WAA attacks, they used model inversion to generate audio adversarial examples. Evolutionary algorithm based black-box attacks is strong, but the number of queries to black-box ASR systems remains high. It is necessary to be optimized.

## V. CONCLUSION

Although researchs on adversarial example attacks against ASR systems have not been around for long, there are quite a few relevant achievements. On white-box attacks, on the one hand, the main work is to find out how to construct universal targeted adversarial attacks. On the other hand it need to be known how to reduce the perception of human ear based on psychoacoustics, and to combine more new methods to try in terms of gradient descent to update the adversarial examples. In terms of black-box attacks, it is to try in the featurization layer; to switch to different evolutionary algorithms to generate adversarial examples or to design an alternative model so that the number of queries to the black-box in the process of constructing the alternative model is as small as possible, and make the alternative model not overfitted in spite of the small amount of training data.

This paper does not focus on the adversarial defense of ASR systems. In fact, there are many processes for audios before featurization, such as VAD, data augmentation, data cleaning, etc. Except training a network to classify audios, those processes above actually contain almost means to deal with adversarial example attacks against ASR systems.

In the future, ASR systems will be used in more and more circumstances. Despite the availability of many AI model defense methods, there are still many shortcomings. We need to solve the corresponding security problems one by one for these attack methods. For example, in terms of the various black-box attacks mentioned in this paper, the genetic algorithm-based adversarial example attack is not quite the same as other adversarial attacks, which traverses almost the entire example space, requiring our system to be robust to the entire example space with strong randomness, just like Web penetration and PWN in various CTF competitions, which requires us to have industrial-grade ASR systems with Mature sweeping vulnerability charter and measures. Therefore, the security of ASR system still has a long way to go.

ACKNOWLEDGMENT

This work is supported in part by the Major Key Project of PCL (Grant No. PCL2022A03), the Guangzhou Science and technology planning project (No. 202102010507), the Key Laboratory of MIIT for Intelligent Products Testing and Reliability 2021 Key Laboratory Open Project Fund (No.CEPRE2022-02), and Guangzhou Higher Education Innovation Group (No. 202032854). Z. Gu is the corresponding author.